\let\accentvec\vec
\documentclass[runningheads,a4paper]{llncs}

\let\vec\accentvec

\usepackage{amssymb}
\usepackage{graphicx}
\usepackage{url}
\usepackage[T1]{fontenc} 
\usepackage{amsmath} 

\usepackage{listings}
\usepackage{algorithmic}
\usepackage{algorithm}
\usepackage{array}
\usepackage{proof}
\usepackage{wrapfig} 

\floatname{algorithm}{Code}
\newcommand{\nc}{\newcommand}
\nc{\algofont}[1]{\sffamily{#1}}   
\nc{\res}[1]{\textsf{#1}}          
\nc{\cod}[1]{\emph{#1}}            
\nc{\ans}{{\mathcal{S}\mathcal{Q}\mathcal{L}}}   
\nc{\select}{\res{select }}                       
\nc{\comment}[1]{}                       

\nc{\multileft}{ \{\!| }
\nc{\multiright}{  |\!\} }
\nc{\multi}[1]{ \multileft {#1} \multiright }
\nc{\evalSQL}[2]{\parallel#1\parallel_{#2}} 
\nc{\mlarge}[1]{\mbox{\Large $#1$}}
\nc{\lsigma}{\mlarge{\sigma}}
\nc{\lrho}{\mlarge{\rho}}
\nc{\lpi}{\mlarge{\Pi}}
\nc{\proofend}{\hfill$\blacksquare$}

\newcommand{\myfontcodesize}{\fontsize{11}{11}}
\newcommand{\mytt}[1]{{\myfontcodesize \texttt{#1}}}

\title{Intuitionistic Logic Programming for SQL \\ 
       (Extended Abstract)
    }
\titlerunning{Intuitionistic Logic Programming for SQL}

\author{Fernando S\'aenz-P\'erez}
\institute{Dept. Ingenier\'{\i}a del Software e Inteligencia Artificial\\
    Declarative Programming Group\\
    Universidad Complutense de Madrid, Spain\\
    \email{fernan@sip.ucm.es} }

\begin{document}

\maketitle

\begin{abstract}
Intuitionistic logic programming provides the notion of embedded implication in rule bodies, which can be used to reason about a current database modified by the antecedent. This can be applied to a system that translates SQL to Datalog to solve SQL \mytt{WITH} queries, for which relations are locally defined and can therefore be understood as added to the current database. In addition, assumptions in SQL queries as either adding or removing data can be modelled in this way as well, which is an interesting feature for decision-support scenarios. This work suggests a way to apply intuitionistic logic programming to SQL, and provides a pointer to a working system implementing this idea.
\end{abstract}

\section{Introduction}

SQL is the {\em de facto} relational database query language that stands still  \cite{Atzeni:2013:RMD:2503792.2503808} despite the advent of new trends as Big Data, NoSQL, RDF stores and others.
It builds upon the Codd's \cite{Codd70,Codd72} seminal relational data model accompanied by an algebra and calculus to operate on data.
Former proposals such as \cite{Grefen1994} better provide a formal framework for current SQL implementations.
As a query language, SQL can be well understood from Codd's tuple relational calculus but also from logic programming (in particular, \cite{Ullman88} includes equivalences between relational operations and logic rules).
However, among other features beyond the original relational model, SQL provides the notion of temporary view defined in {\tt WITH} clauses (as described in Section \ref{with}), whose definition is available only to the query in which it occurs \cite{Silberschatz6th}.
This is no longer representable either in relational formal languages or directly in logic programming.\footnote{Obviously, logic programming implementations provide general-purpose languages as Prolog that can emulate a temporary definition.}

Here is when intuitionistic logic programming may come at help to providing first-class citizen semantics:
Approaches as \cite{DBLP:journals/jlp/Gabbay85,DBLP:journals/jlp/McCarty88,MillerModulesJLP,Bonner90hypotheticaldatalog,Hodas94} fit into this logic, an extension of logic programming including in particular embedded implications.
Adding negation to intuitionistic logic programming might develop paradoxes which are circumvented in \cite{bonner90adding} by dealing with two kind of implications: for rules ($\leftarrow$) and for goals ($\Leftarrow$, i.e., an embedded implication).
Whereas in the formula $A \leftarrow B$, the atom $B$ is ''executed" for proving $A$, in the formula $A \Leftarrow B$, the atom $B$ is ``assumed" to be true for proving $A$. 
Hypothetical Datalog \cite{Bonner89hypotheticaldatalog,Bonner90hypotheticaldatalog,bonner90adding,bonner89expressing} incorporated this logic and has been a proposal thoroughly studied from semantic and complexity point-of-views.
The work \cite{sae13c-ictai13} (recalled in Section \ref{hypothetical-datalog}) presented an extended (w.r.t. \cite{Bonner89hypotheticaldatalog,Bonner90hypotheticaldatalog,bonner90adding,bonner89expressing}) intuitionistic setting along with an implementation in the deductive system DES \cite{saenzDES}, in which a rule is accepted in the antecedent of an embedded implication, and not only facts as in \cite{bonner90adding}.

Driven from the need for supporting a broader subset of SQL in this system, we show how to take advantage of the intuitionistic embedded implication to model {\tt WITH} SQL queries in a logic setting, an application which has not been proposed to the best of our knowledge so far.
Thus, as shown in Section \ref{translating}, it is possible to have such SQL queries translated into Datalog, and can be therefore processed by a deductive engine. 
But Hypothetical Datalog is powerful enough to even apply the same technique to model assumptions in SQL queries, with the (non-standard) clause \mytt{ASSUME}.
This clause enables both positive and negative assumptions on data, as shown in Section \ref{assume}, which are useful for modelling ``what-if'' scenarios.
Finally, we present the deductive system DES at work with examples of \mytt{WITH} and \mytt{ASSUME} queries in Section \ref{examples}.
Our approach is also useful for connecting with external relational database systems which cannot process these clauses.
DES then behaves as a front-end capable of processing either novel or unsupported features in such systems.

\section{The SQL {\tt WITH} Clause}
\label{with}

Typically, complex SQL queries are broken-down for applying the {\em divide-and-conquer} principle as well as for enhancing readability and maintenance.
Introducing intermediate views with {\tt CREATE VIEW} statements is in the order of the day, but this might neither be recommendable (making these views observable for other users) nor possible (only certain users with administration permissions are allowed to create views). 
The {\tt WITH} clause provides a form of encapsulation in SQL by locally defining those broken-down views, making their realms to pertain to the context of a given query.
Next, the syntax of a query $Q$ including this clause is recalled:

\vspace{2mm}
\begin{tabular}{ll}
\mytt{WITH} &\mytt{{\em R}$_1$ AS {\em SQL}$_1$,}\\
            &\mytt{\ldots,} \\
            &\mytt{{\em R}$_n$ AS {\em SQL}$_n$}\\
\mytt{{\em SQL}}
\end{tabular}

\noindent where each \mytt{{\em R}$_i$} is a temporary view name defined by the SQL statement \mytt{{\em SQL}$_i$}, and which can be referenced only in \mytt{{\em SQL}}, the ultimate query that builds the outcome of the query $Q$.
This query can be understood as a relation with name \mytt{{\em R}} and defined with the DDL statement \mytt{CREATE VIEW {\em R} AS} $Q$. 

With respect to the semantics of an SQL query, we recall and adapt the notation in \cite{cgs12a-flops2012} which in turn is based on \cite{Grefen1994}.
A {\em table instance} is a multiset of facts (following logic programming instead of relational databases).
A {\em database instance} $\Delta$ of a database
schema is a set of table instances, one for each defined table ({\em extensional} relation) in the database.
The notation $\Delta(T)$ represents the instance of a table $T$ in $\Delta$.
Each query or view ({\em intensional} relation) $R$ is defined as a multiset of facts,
and $\Phi_R$ represents the ERA (Extended Relational Algebra) expression associated to an SQL query or view $R$, as explained in \cite{Molina08}.
An intensional relation usually depends on previously defined relations, and sometimes it will be useful to write  $\Phi_R(R_1,\dots,R_n)$ indicating that $R$ depends on $R_1, \dots, R_n$.
Here, we assume that each extensional relation in a database instance has attached type information for each one of its named arguments.
As well, each intensional relation argument receives its type via inferencing and arbitrary names if not provided in its definition.
Tables are denoted by their names, that is, $\Phi_T = T$ if $T$ is a table.
	\begin{definition}  \label{def:SQLERA} {\em
	The \emph{computed answer} of an ERA expression $\Phi_R$ with respect to some schema instance $\Delta$ is denoted by $\evalSQL{\Phi_R}{\Delta}$, where:
	\begin{itemize}
		\item If $R$ is an extensional relation, $\evalSQL{\Phi_R}{\Delta} =\, \Delta(R)$.
		\item If $R$ is an intensional relation and $R_1, \dots, R_n$ the relations defined in $R$, then
		$\evalSQL{\Phi_R}{\Delta} =\, \Phi_R(\evalSQL{\Phi_{R_1}}{\Delta}, \dots, \evalSQL{\Phi_{R_n}}{\Delta})$. \hfill $_\square$
	\end{itemize}
}
    \end{definition}

Queries are executed by SQL systems. 
The answer for a query $Q$ and a database instance $\Delta$ in an implementation is represented by $\ans_\Delta$($Q$). The notation $\ans_\Delta$($R$) abbreviates $\ans_\Delta$(\mytt{SELECT * FROM $R$}). 
In particular, we assume the existence of \emph{correct} SQL implementations.

	\begin{definition} \label{def:correctSQL} {\em 
	A \emph{correct} SQL implementation verifies that $\ans_\Delta$($Q$) = $\evalSQL{\Phi_Q}{\Delta}$ for every query $Q$.\hfill $_\square$
}
	\end{definition}



\section{Hypothetical Datalog}
\label{hypothetical-datalog}

Hypothetical Datalog is an extension of function-free Horn logic 
\cite{bonner90adding}.
%
Following \cite{sae13c-ictai13}, the syntax of the logic is first order and includes a universe of constant symbols, a set of variables and a set of predicate symbols.
For concrete symbols, we write variables starting with either an upper-case letter or an underscore, and the rest of symbols starting with lower-case.
Removing function symbols from the logic is a condition for finiteness of answers, a natural requirement of relational database users.
As in Horn-logic, a rule has the form $A \leftarrow \phi$, where $A$ is an atom and $\phi$ is a conjunction of goals.
Since we consider a hypothetical system,
a goal can also take the form $R_1 \land \ldots \land R_n \Rightarrow G$, a construction known as an {\em embedded implication}.
The following definition captures the syntax of the language, where $vars(T)$ is the set of variables occurring in $T$:

\begin{definition} {\em 
	\quad \\
	$R := A \mid A \leftarrow G_1 \land \ldots \land G_n$ \\
	$G := A \mid \neg G \mid R_1 \land \ldots \land R_n \Rightarrow G$\\
	\noindent where $R$ and $R_i$ stand for rules, $G$ and $G_i$ for goals, $A$ for an atom  (possibly containing variables and constants, but no compound terms), and $\bigcup vars(R_i) \cap vars(R) = \emptyset$, and $vars(R_i)$ and $vars(G)$ are disjoint.\hfill $_\square$
}
\end{definition}

Disjoint conditions ensure that assumed rules do not depend on actual substitutions along inference, i.e., assumed rules take the form they have in the program.

Semantics is built with a stratified inference system which can be consulted in \cite{sae13c-ictai13}.
Here we recall the inference rule for the embedded implication:\footnote{Each rule in this inference system is read as: If the formulas above the line can be inferred, then those below the line can also be inferred.}
For any goal $\phi$ and database instance $\Delta$:

\begin{center}
\medskip
$\infer[]{\Delta \vdash R_1 \land \ldots \land R_n \Rightarrow \phi}{\Delta \cup \{R_1, \ldots, R_n\} \vdash \phi}$
\medskip
\end{center}

This means that for proving the conclusion $\phi$, rules $R_i$, together with the current database instance $\Delta$ can be used in subsequent inference steps.
The {\em unified stratified semantics} defined in \cite{sae13c-ictai13} builds a set of axioms $\mathcal{E}$ that provides a means to assign a meaning to a goal as: 
$solve(\phi,\mathcal{E}) = \{\Delta \vdash id:\psi \in \mathcal{E} ~ \mbox{such that} ~ \phi\theta = \psi\}$, where $\theta$ is a substitution and each axiom in $\mathcal{E}$ is mapped to the database $\Delta$ it was deduced for, and the inferred fact $\psi$ is labelled with its data source (for supporting duplicates).
We use $\Delta(\mathcal{E})$ to denote the multiset of facts $\psi$ so that $\Delta \vdash id:\psi \in \mathcal{E}$ for any $id$.
So, this inference rule captures what SQL {\tt WITH} statements need if translated to Hypothetical Datalog, because each $R_i$ can represent each temporary view definition, as will be shown in the next section.

\section{Translating SQL into Datalog}
\label{translating}

We consider standard SQL as found in many textbooks (e.g., \cite{Silberschatz6th}), but also allowing \verb|FROM|-less statements, i.e., providing a single-row output constructed with the comma-separated expressions after the \verb|SELECT| keyword (Oracle, for instance, resorts to feed the row from the \verb|dual| table to express the same feature).
Here, we define a function \mytt{{\em SQL\_to\_DL}} that takes a relation name  and an SQL statement as input and returns a multiset of Datalog rules providing the same meaning as the SQL relation for a corresponding predicate with the same name as the relation.
The following (incomplete) definition for this function includes only a couple of the basic cases, where others can be easily developed from \cite{Ullman88}.
From here on, set-related operators and symbols refer to multisets, as SQL relations can contain duplicates. 

	 
\medskip
\noindent {\em \% Basic SELECT statement} \\
\noindent \mytt{{\em SQL\_to\_DL}}($r$, \mytt{SELECT A$_1$,\ldots,A$_n$ FROM $Rel$ WHERE $Cond$}) =\\
\indent \{ $r(\overline{X_i}) \leftarrow DLRel(\overline{X_i}), DLCond(\overline{X_j}) ~ \} \bigcup RelRules \bigcup CondRules$, \\
where \mytt{{\em SQLREL\_to\_DL}}$(Rel)=(DLRel(\overline{X_i}), RelRules)$, and \\
\indent \mytt{{\em SQLCOND\_to\_DL}}$(Cond)=(DLCond(\overline{X_j}), CondRules)$
\medskip

\noindent {\em \% Duplicate-preserving union} \\
\noindent \mytt{{\em SQL\_to\_DL}}($r$, \mytt{{\em SQL}}$_1$ \mytt{UNION ALL} \mytt{{\em SQL}}$_2$) = \\
\indent 
\mytt{{\em SQL\_to\_DL}}($r$, \mytt{{\em SQL}}$_1$) $\bigcup$
\mytt{{\em SQL\_to\_DL}}($r$, \mytt{{\em SQL}}$_2$) 
\medskip


Here, each \mytt{A}$_i$ is an argument name present in the relation $Rel$ with corresponding logic variable $X_i$. 
$Rel$ is constructed with either a single defined relation (table or view), or a join of relations, or an SQL statement.
Function \mytt{{\em SQLREL\_to\_DL}} (resp., \mytt{{\em SQLCOND\_to\_DL}}) takes an SQL relation (resp. condition) and returns a goal and, possibly, additional rules which result from the translation.
Variables $\overline{X_j}$ come as a result of the translation of the condition $DLCond$ to a goal. 
As well, some basic cases are presented next for these functions, where $GoalName$ is an arbitrary, fresh new goal name:

\medskip
\noindent {\em \% Extensional/Intensional Relation Name} \\
\noindent \mytt{{\em SQLREL\_to\_DL}}$(RelName)=(RelName(\overline{X_i}), \{ \})$ \\
\noindent where $\overline{X_i}$ are the $n$ variables corresponding to the $n$-degree relation $RelName$.
\medskip

\noindent {\em \% SQL Statement} \\
\noindent \mytt{{\em SQLREL\_to\_DL}}$(SQL)=(GoalName(\overline{X_i})$, \mytt{{\em SQL\_to\_DL}}$(GoalName, SQL)$)\\
\noindent where $\overline{X_i}$ are the $n$ variables corresponding to the $n$-degree statement $SQL$.
\medskip

\noindent {\em \% NOT IN Condition} \\
\noindent \mytt{{\em SQLCOND\_to\_DL}}(\mytt{A$_i$ NOT IN} $Rel)=(not ~ DLRel(\overline{X_j}), RelRules)$\\
where \mytt{{\em SQLREL\_to\_DL}}$(Rel)=(DLRel(\overline{X_j}), RelRules)$, and $X_i \in \overline{X_j}$ is the corresponding variable to argument \mytt{A}$_i$.
\medskip

Completing this function by including the \mytt{WITH} statement is straightforward because every temporary view can be represented by a predicate resulting from the translation of the temporary view definition into Datalog rules.
Assuming such predicates as the antecedent of an embedded implication can be used to augment the (local, temporary) database for interpreting the meaning of the translated SQL outcome query:

\medskip
\noindent
	{\mytt{{\em SQL\_to\_DL}}($r$, \mytt{WITH $r_1$ AS {\em SQL}$_1$, \ldots, $r_n$ AS {\em SQL}$_n$ {\em SQL}}) =}\\
\indent	\{ $r(\overline{X_i}) \leftarrow$ \\
\indent	\hspace*{1.7mm}
	$\land($\mytt{{\em SQL\_to\_DL}}($r_1$,\mytt{{\em SQL}}$_1$)) $\land$ \ldots \\
\indent	\hspace*{1.7mm}
	$\land($\mytt{{\em SQL\_to\_DL}}($r_n$,\mytt{{\em SQL}}$_n$)) $\Rightarrow$ 
	$s(\overline{X_i})$ \} $\bigcup$ \texttt{{\em SQL\_to\_DL}}($s$,\texttt{{\em SQL}})  
\medskip

\noindent where $\land(Bag)$ denotes $B_1 \land \cdots \land B_m$ ($B_i \in Bag$).

The following theorem establishes the semantic equivalence of an SQL relation and its counterpart Datalog translation.

\begin{theorem} {\em
	The semantics of an SQL $n$-degree relation $r$ defined by the query $Q$ on a database instance $\Delta$ coincides with the meaning of a goal $r(\overline{X_i})$, $1 \leq i \leq n$, for $\Delta' = \Delta \bigcup$ \mytt{{\em SQL\_to\_DL}}($r$,$Q$), that is:
	$\ans_\Delta(Q) = \Delta(solve(r(\overline{X_i}),\mathcal{E}))$, where $\mathcal{E}$ is the unified stratified semantics for $\Delta'$.}  \hfill $_\square$
\end{theorem}

\section{Beyond the WITH Clause: Expressing Assumptions}
\label{assume}

As a novel feature, hypothetical SQL queries (absent in the standard) were introduced (inspired in \cite{nss-flops08}) in DES version 2.6 for solving ``what-if'' scenarios.
Syntax for such queries is:

{\myfontcodesize
	\vspace{3mm}
	\noindent {\tt ASSUME {\em SQL}$_1$ IN {\tt {\em Rel}}$_1$, ..., {\em SQL}$_n$ IN {\tt {\em Rel}}$_n$}
	{\tt {\em SQL};}
	\vspace{3mm}
}

\noindent which makes to {\em assume} the result of {\tt {\em SQL}}$_i$ in {\tt {\em Rel}}$_i$ when processing {\tt {\em SQL}}.
This means that the semantics of each {\tt {\em Rel}}$_i$ is either overloaded (if the relation already exists) or otherwise defined with the facts of {\tt {\em SQL}}$_i$.
Implementing this resorted to {\em globally} define each {\tt {\em Rel}}$_i$, which is not the expected behaviour as its definition must be local to {\tt {\em SQL}}.
Roughly, solving an \mytt{ASSUME} query resorted to overload the meanings of each {\tt {\em Rel}}$_i$ (by inserting the required facts) before computing {\tt {\em SQL}} and, after solving, to restore them (by deleting the same facts).
This also precluded nested assumptions, and such statements were allowed only as top-level queries but not as part of query definitions.
For instance, if it would be allowed, the following query would be incorrectly computed in that scenario:

{\myfontcodesize
	\begin{verbatim}
ASSUME SELECT 1 IN r(a), 
       (ASSUME SELECT 2 IN r(a) SELECT * FROM r) IN s 
SELECT * FROM r,s;
	\end{verbatim}
}

\noindent because the meaning of \mytt{r} in the context of \mytt{SELECT * FROM r,s} would be overloaded with both \mytt{\{(1)\}} and \mytt{\{(2)\}}, instead of just with \mytt{\{(1)\}}.

Applying hypothetical reasoning in this case solves this issue, allowing us not only to use nested assumptions in both top-level queries and views, but also to take advantage of negative assumptions.
A negative assumption allows to {\em remove} facts from the meaning of a relation, which broadens the applicability of queries in decision-support scenarios.
To specify negative assumptions, \mytt{NOT IN} is used instead of just \mytt{IN}.
Hypothetical Datalog in \cite{Sae15c} introduces the notion of {\em restricted predicate} to handle negative assumptions in embedded implications. 
A restricted predicate includes at least a restricting rule whose head is an atom preceded by a minus sign.
Its meaning is the set of facts deduced from regular rules minus the set of facts deduced from restricting rules.
So, a negative assumption is modelled with a restricting rule in the antecedent of an embedded implication, so that the translation from SQL to Hypothetical Datalog for \mytt{ASSUME} statements becomes:

\medskip
\noindent
\begin{tabular}{m{0.1cm}m{0.1cm}l}
	\multicolumn{3}{l}{\mytt{{\em SQL\_to\_DL}}($r$, \mytt{ASSUME {\em SQL}$_1$ [NOT] IN $r_1$,\ldots,{\em SQL}$_n$ [NOT] IN $r_n$} \mytt{{\em SQL}}) =}\\
	& \{ & $r(\overline{X_i}) \leftarrow$ \\
	& & $\land$(\mytt{{\em SQL\_to\_DL}}($r_1$, \mytt{{\em SQL}}$_1$)$[$\mytt{[-]}$r_1$/$r_1]$) $\land$
	\ldots \\
	& & $\land$(\mytt{{\em SQL\_to\_DL}}($r_n$, \mytt{{\em SQL}}$_n$)$[$\mytt{[-]}$r_n$/$r_n]$) \texttt{=>} 
	$s(\overline{X_i})$ \} \\
	& & $\cup$ $\texttt{{\em SQL\_to\_DL}}(s,\texttt{{\em SQL}})$
\end{tabular}
\medskip

\noindent where $A[B/C]$ represents the application of the syntactic substitution $C$ by $B$ in all the rule heads in $A$, \mytt{[{\em T}]} represents that \mytt{{\em T}} is optional but it must occur if the corresponding $i$-th entry also occurs (i.e., if \mytt{NOT} occurs in the assumption for $r_i$, then \mytt{-} also occurs in the corresponding substitution).

\section{Playing with the System}
\label{examples}

Translating an SQL query to Datalog in a practical system involves more features that the ones briefly suggested before and are out of the scope of this extended abstract.
For example, the {\tt SELECT} list can include expressions and scalar SQL statements, nested statements can be correlated, aggregate functions and grouping can be used, and so on.
Also, from a capacity point-of-view, a needed stage in the translation is folding/unfolding of rules \cite{prolog} to simplify the Datalog program resulting from the translation.
This is quite relevant because deducing the meaning (either complete or restricted to a given call) of the involved relations along query solving is needed, therefore significantly augmenting their space and time requirements.
Next, we introduce a couple of examples of this translation with the system DES \cite{saenzDES}, which in particular supports such features and inputs from several query languages, including Datalog and SQL.
Here, we resort to the actual textual syntax of Datalog rules in this system, which follows the syntax of Prolog.

\begin{wrapfigure}{r}{45mm}
	\vspace{-35pt}
	\begin{center}
		\begin{tabular}{cc}
			\begin{minipage}[c]{2cm}
				\begin{tabular}{|c|}
					\hline 
					\mytt{student}\\
					\hline 
					\mytt{(adam)} \\ 
					\mytt{(bob)} \\ 
					\mytt{(pete)} \\ 
					\mytt{(scott)} \\ 
					\hline 
				\end{tabular}
			\end{minipage}
			& 
			\begin{tabular}{|c|}
				\hline 
				\mytt{take}\\
				\hline 
				\mytt{(adam,db)} \\ 
				\mytt{(pete,db)} \\ 
				\mytt{(pete,lp)} \\ 
				\mytt{(scott,lp)} \\ 
				\hline 
			\end{tabular}	
		\end{tabular}
	\end{center}
	\vspace{-25pt}
\end{wrapfigure}
Let us consider a database containing the relations \mytt{student(name)} and \mytt{take(name, title)}.
The first one states names of students and the second one the course (\mytt{title}) each student (\mytt{name}) is enrolled in. 
Types can be specified either with a Datalog assertion (as \mytt{:-type(student(name:string))} for the first case) or a DDL SQL statement (as \mytt{create table take(name string, title string)} for the second one, where a foreign key \mytt{take.name} $\rightarrow$ \mytt{student.name} could be stated as well). 
We consider the database instance depicted in the tables above.
%


The next SQL statement (looking for students that have not been already enrolled in a course)
is translated as follows in a system session with DES 4.0:

{\myfontcodesize
	\begin{verbatim}
	DES> select * from student where name not in 
        (select name from take)
	Info: SQL statement compiled to:
	  answer(A) :- student(A), not take(A,_B).
	answer(student.name:string) ->
	{ answer(bob) }
	Info: 1 tuple computed.          
	\end{verbatim}
}

This example shows a few of things.
First, as a query $Q$ is allowed at the system prompt, the call to the translation function becomes \mytt{{\em SQL\_to\_DL}}(\mytt{answer},$Q$), i.e., the outcome relation is automatically renamed to the reserved keyword \mytt{answer}.
Second, the outcome schema \mytt{answer(student.name:string)} shows that the single output argument comes from the argument \mytt{name} of the relation \mytt{student}, with type \mytt{string}.
Third, following the definition of the translation function, this query should be translated into:

{\myfontcodesize
	\begin{verbatim}
	answer(A) :- student(A), goal1(A).
	goal1(A) :- not take(A,_B)
	\end{verbatim}
}

But folding/unfolding simplifies this as it was displayed in the system session.
These translations have been displayed because it was specified so by issuing the command \mytt{/show\_compilations on}. 
Finally, non-relevant variables to a rule outcome are underscored (otherwise, they would be signalled as anonymous).
This is important in this case to identify as safe the rule in which this underscored variable occurs.
Classical safety \cite{Ullman88} would tag the rule \mytt{answer(A) :- student(A), not take(A,\_B)} as unsafe, but an equivalent set of safe rules can be found: \mytt{answer(A) :- student(A), not goal1(A)} and \mytt{goal1(A) :- take(A,\_B)}.
Underscored variables are a means to encapsulate this form of safety, which is identified by the system and processed correspondingly.
In general, there can be several rules for \mytt{answer} (e.g., when a \mytt{UNION} is involved) and others on which this predicate depends on.

As an example of a \mytt{WITH} query, the following statement defines the relation \mytt{grad} intended to retrieve the eligible students for graduation (those that took both \mytt{db} and \mytt{lp} in this tiny example):

{\myfontcodesize
	\begin{verbatim}
	DES> with grad(name) as 
	       (select student.name
	        from student, take t1, take t2
	        where student.name=t1.name
	          and t1.name=t2.name 
	          and t1.title='db' and t2.title='lp') 
	     select * from grad;
	Info: SQL statement compiled to:
	  answer(A) :-
	  (grad(B) :- student(B), take(B,db), take(B,lp))
	  =>
	  grad(A).
	answer(grad.name:varchar(30)) ->
	{ answer(pete) }
	Info: 1 tuple computed.          
	\end{verbatim}
}

As an example of an \mytt{ASSUME} query, we reuse the \mytt{grad} definition above, assume that \mytt{adam} is not an eligible student, and that  \mytt{adam} and \mytt{scott} took \mytt{lp} and \mytt{db} respectively:

{\myfontcodesize
	\begin{verbatim}
DES> assume 
       (select 'adam') not in student, 
       (select 'adam','lp' union all select 'scott','db')
         in take, 
       (select student.name from student, take t1, take t2
         where student.name=t1.name and t1.name=t2.name and
               t1.title='lp' and t2.title='db') in grad(name) 
     select * from grad;
Info: SQL statement compiled to:
  answer(A) :-
  -student(adam) /\ take(adam,lp) /\  take(scott,db) /\
  (grad(B) :- student(B), take(B,lp), take(B,db))
  =>
  grad(A).
answer(grad.name:varchar(30)) ->
{ answer(pete), answer(scott) }
Info: 2 tuples computed.          
\end{verbatim}
}

Here, the assumption on \mytt{student} is negative and is compiled to a restricting fact. 
The second one is compiled to a couple of facts because of the union.
The last one is the same as the previous example.
The SQL statement after the assumptions simply leads to the goal \mytt{grad(A)}, for which even when \mytt{adam} took the courses to graduate, he was removed as an eligible student and therefore from the answer.

If the extensional relations \mytt{student} and \mytt{take} are already defined in an external relational database (as, e.g., MySQL or PostgreSQL), they can be made available to DES via an ODBC connection (with the command \mytt{/open\_db}), and queried as if they were local \cite{Sae14a}.
This way, DES behaves as a front-end for both straight calls to native (i.e., supported by the external relational system) SQL queries and non-native queries (as those including \mytt{ASSUME}).
For non-native statements, prepending the command \mytt{/des} to the query makes DES to handle such queries which are unsupported in the external database.
For example:

{\myfontcodesize
	\begin{verbatim}
	DES> /open_db postgresql 
	DES> /des assume ...
	DES> /open_db mysql 
	DES> /des with ...
	\end{verbatim}
}
  
\noindent obtaining the same answers as before for the same queries (both omitted here in the ellipses).
Note that, in particular, \mytt{WITH} is unsupported in both MySQL and MS Access.

Even when \mytt{WITH} is supported in several relational database systems, they are somewhat restricted because, referring to the syntax in Section \ref{with}, \mytt{{\em SQL}} cannot contain a \mytt{WITH} clause, whereas we do allow for it.

\section{Conclusions}
This work has presented a proposal to take advantage of intuitionistic logic programming to model both temporary definitions (with the \mytt{WITH} clause) and assumptions (with the \mytt{ASSUME} clause) in SQL.
Its motivation lies in providing a clean semantics that makes assumptions to behave as first-class citizen in the object language.
The deductive database system DES was used as a test bed to experiment with assumptions, translating SQL queries into Hypothetical Datalog.
Further, this system can be used as a front-end to relational systems lacking features as the \mytt{WITH} clause.
The most related work is \cite{ANSS13d}, which includes assumptions in SQL with a tailored semantics, and generates SQL scripts implementing fixpoint computations.
With respect to the intuitionistic formal framework, our work is based on \cite{Bonner89hypotheticaldatalog,Bonner90hypotheticaldatalog,bonner90adding,bonner89expressing} and adapted to assume rules and deal with duplicates in \cite{sae13c-ictai13}.
However, it is not powerful enough to include embedded universal quantifiers in premises as in \cite{bonner89expressing}, which provides the ability to create new constant symbols hypothetically along inference.
Though this is not directly applicable to the current work, it is indeed an interesting subject to explore by considering that domains can be finitely constrained in practical applications, as with foreign keys.

\section*{Acknowledgements}
Thanks to the anonymous referees for their suggestions to improve this work, which has been partially supported by the Spanish MINECO project CAVI-ART (TIN2013-44742-C4-3-R), Madrid regional project N-GREENS Software-CM (S2013/ICE-2731)  and UCM grant GR3/14-910502.


\end{document}